\documentclass[pra,aps,10pt, twocolumn, floatfix]{revtex4-1} 

\usepackage {graphicx} 
\usepackage {amssymb} 
\usepackage {amsmath}

\newcommand{\reffig}[1]{Fig.~\ref{#1}}
\newcommand{\refeq}[1]{Eq.~(\ref{#1})}
\newcommand{\refeqs}[2]{Eqs.~(\ref{#1})-(\ref{#2})}

\newcommand{\vect}[1]{\mathrm{\mathbf{#1}}} 

\begin{document}

\title{Transient Cherenkov radiation from an inhomogeneous string
  excited by an ultrashort laser pulse at superluminal velocity}

\author{R.M. Arkhipov} \affiliation{Weierstrass Institute for Applied
  Analysis and Stochastics Leibniz Institute in Forschungsverbund
  Berlin e. V., Mohrenstr. 39 10117 Berlin Germany, Faculty of
  Physics, St. Petersburg State University, Ulyanovskaya 1,
  Petrodvoretz, St. Petersburg 198504 Russia} \author{I. Babushkin}
\affiliation{Institute of Mathematics, Humboldt University of Berlin,
  Rudower Chaussee 25 12489 Berlin Germany} \author{
  M.K. Lebedev, Yu.A. Tolmachev, M.V. Arkhipov} \affiliation{Faculty of Physics,
  St. Petersburg State University, Ulyanovskaya 1, Petrodvoretz,
  St. Petersburg 198504 Russia }

\date{\today}

\begin{abstract}
  An optical response of one-dimensional string made of dipoles with a
  periodically varying density excited by a spot of light moving along
  the string at the superluminal (sub-luminal) velocity is
  theoretically studied.
  The Cherenkov radiation in such system is rather
    unusual, possessing both transient and resonant character. We show
    that under certain conditions, in addition to the resonant
    Cherenkov peak another Doppler-like frequency appears in the
    radiation spectrum. Both linear (small-signal) and
    nonlinear regimes as well as different string topologies are
    considered.
\end{abstract}


\maketitle

\section{\label{sec:section-1}Introduction}

The problem of superluminal motion and its
existence in nature attracts attention of
various researchers for rather long time. At the turn of XIX-XX
centuries, O. Heaviside and A. Sommerfeld considered radiation of
charged particles moving in vacuum at the velocity greater than the
velocity of light in vacuum $c$ (see
\citep{landau84:book,jelley58:book,bolotovskii62,bolotovskii90,recami00,recami01,
  recami03,bolotovskii05a,kobzev10,malykin12} and references
therein). However, their works were forgotten for many years because the
special theory of relativity "bans" such motions. Further analysis has
shown that the "prohibited" are only those motions that involve signal
(information) transfer at the superluminal velocity, this strong
prohibition being related to the violation of the causality principle
~\cite{landau84:book,jelley58:book,kirzhnits74,kobzev10}.

If a charged particle moves faster than light in some medium
so called Cherenkov radiation occurs. It 
is emitted typically
into a cone with the angle depending on the ratio of the particle
velocity and the speed of light in the media. Similar conical emission
can appear also in nonlinear optical parametric processes
\cite{you91,you92,chalupczak94,paul02,vaiifmmode07}.  Not only
particles but also spots of light can propagate faster than the phase
velocity of light in particular medium
\cite{bolotovskii05a,mironenko99,mcdonald00,doilnitsyna06,chen11a,arkhipov12}. Those
can be optical pulses and solitons in fibers or filaments
\cite{wai90,karpman93,karpman93a,akhmediev95,demircan11,
  demircan12,driben12,yulin13} as well as in other optical systems
\cite{luo03,bakunov05,damico07,hu07,bakunov07,babushkin07,yao12,burlak12,liu12,fernandes12}.
Cherenkov radiation in various periodically modulated media,
with modulations both in
space~\cite{smith53,luo03,longhi03} and time
~\cite{zurita-sanchez12} was also considered.  The intersection
point of two wavefronts can also move at the velocity exceeding that
of light ~\cite{bolotovskii05a,salo00}.  Similar situation occurs when
a short plane-wave pulse crosses a flat screen (or the plane
diffraction grating) ~\cite{smith53,bolotovskii05a}. In this case, the
intersection of the pulse and the screen moves along the screen at the
velocity $V=c/\sin\beta > c$ (here $\beta$ is the angle of wave
incidence) ~\cite{frank42, bolotovskii72}.

Despite of the great number of 
various configurations studied in the
context of Cherenkov radiation, in all those cases such radiation has
the same nature. Namely, it is a result of
interference of the secondary waves emitted by the
moving ``particle''.

The temporal shape of the radiating wave and thus the spectrum of
radiation can be significantly different depending on the particular
situation. For instance, the spectrum of radiation induced by a
charged particle moving faster than the phase velocity of light is
rather unstructured \cite{landau84:book,jelley58:book}. In many other cases,
resonances may occur \cite{smith53,bolotovskii05a}.  One important
example is the so called Purcell-Smith radiation
\cite{smith53,woods95} appearing as a charged particle moves in the vicinity
of a periodic structure. Also, a moving and at the same time
oscillating dipole emits the Cherenkov radiation characterized with
well defined resonance \cite{yao12}, similar situation is realized for
optical solitons propagation \cite{skryabin10}.

In the present article, we consider in details the Cherenkov-type
radiation in the case of a one-dimensional (1D) string formed of two-level atoms with a
spatially-periodic modulated number density. This system is excited at
the superluminal (sub-luminal) velocity at the point of intersection
of the string with a moving spot of light. This geometry is imposed by
recent advances in optical technologies which allow for the reliable
control of matter properties on the spatial level of the order of
wavelength of light or even much smaller that allows to create
quasi-1D objects (see \cite{novotny11,biagioni12} and references
therein). Although our consideration is rather general, we bear in
mind the spatial properties nanoantenna or quantum dots arrays
\cite{novotny11,biagioni12,sun13,feuillet-palma13,hartmann99,felici09,mohan10,juska13}
as well as thin microcapillaries. Similar geometry was
recently realized experimentally in \cite{chen11a}.

As we show, in the system considered here the Cherenkov emission have
somewhat unusual character. It possesses a narrow-band
spectrum, with the central frequency at the resonance of dipoles the
string comprises of.  In
  presence of inhomogeneities
of the dipoles density a second Doppler-like frequency appears in the
spectrum even in a linear regime, that is, when the pump remains
weak. We show that this effect does not depend on the string geometry
by considering both a straight and a circularly-shaped string.  With
increasing the pump, when the nonlinear response of two level atoms
becomes important, this new frequency may even significantly overcome
the resonant one.

The structure of the paper is as following: in the Section II
the system is described and  the possible physical
realizations are considered;
Section III describes a linear response of the string
  whereas in Sec. IV the nonlinear dynamics is considered. Concluding remarks are presented in Sec. V.

\section{Physical considerations}

\begin{figure}[h]
\center{\includegraphics[width=\linewidth]{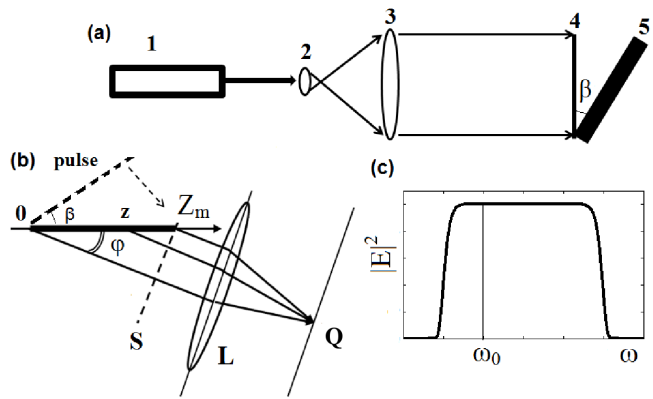}}
\caption{ (a) Excitation of a string at superluminal velocity. 1 - a
  short spectrally broadband laser pulse source, 2,3 -lenses, 4 the
  plane pulse wave.  The intersection of the plane pulse and the medium
  moves along the string at the velocity $V=c/\sin\beta > c$. (b) The
  observation geometry of the string of the length $Z_m$ emission. The observer is placed far away from the string or in the focus of a lens L which collects the
  radiation of the string parallel to the $z$-axis. (c) The source must produce significantly broadband and
  flat spectrum \cite{reimann07, thomson07,kim07,kim08b, babushkin10,
    babushkin10a,koehler11a,babushkin11,berge13}, which includes also
  the resonance frequency $\omega_0$ of the dipoles forming the
  string.}
\label{fig1}
\end{figure}

  The geometry we would like to consider is illustrated in
  \reffig{fig1}(a). A short spectrally broadband optical plane wave
  pulse is emitted by a source 1, passing through lenses 2 and 3 which
  makes the pulse spatially extended. The source must produce
  significantly broadband and flat spectrum \cite{reimann07,
    thomson07,kim07,kim08b, babushkin10,
    babushkin10a,koehler11a,babushkin11,berge13}, which includes also
  the resonance frequency $\omega_0$ of the dipoles forming the
  string.  This spatially extended short in time and in the axial
direction pulse has the form of thin ``sheet of light'' 4, which
illuminates at the angle $\beta$ the string medium parallel to
z-axis. Similar geometry was recently realized experimentally
\cite{chen11a}.

We also assume that the string consists of oscillators (dipoles) with
the resonance frequency $\omega_0$ and decay constant $\gamma$, which
number density $N(z)$ varies periodically along the
string with the spatial period $\Lambda_z$.

  Moreover, we assume that the string is thin: its
thickness is less than the wavelength of light corresponding to the
  resonance frequency $\omega_0$. Such quasi-1D geometry of the system
  suggests that, if the irradiation and observation angles $\beta$ and
  $\varphi$ are not zero (cf. \reffig{fig1}(a)), the secondary radiation
  emitted by any dipole will never hit another dipole on its way to
  the observer.
  It should be noted that even
  for a string which thickness is less than the light wavelength (at
  resonant frequency) not only linear but also nonlinear response is
  possible, leading to optical bistability as well as nontrivial
  dynamical regimes such as pulsations or chaos
  \cite{hupf84,benedict91,logvin92,oraevsky94,loiko96,babushkin98,
    babushkin00,babushkin00a,klyukanov01,paulau04}.

The electric field created by this excitation observed at the remote
position Q is determined by the solution of the wave equation $\square
\vect E = \mu_0\partial_{tt}\vect P$, where $\square
= \partial_{xx}+ \partial_{yy}+ \partial_{zz}-1/c^2 \partial_{tt}$ is
the d'Alembert operator, $c$ is the velocity of light in vacuum. In
particular, if the source is a single dipole,
 that is, $\vect P\propto
\delta(\vect r)$, the observed secondary
  emission at the point $\vect r'$ is:
\begin{equation}
\vect E(\vect r',t) \propto  \partial_{tt} \vect{P}(\vect r, t-|\vect r - \vect
r'|/c). \label{eq:3}
\end{equation}
In the following, we normalize the later relation in such a way that
the coefficient of proportionality between $\vect E$ and
$\omega_0^2\vect P$ is one.

Under those circumstances, the response of dipoles to
  some excitation $\vect E_e(t)$ as seen from the point Q is described by
  the sum of the responses of separate dipoles over the whole string,
  taking into account that the response comes to the point Q delayed
  in time. If the response of a single dipole to the excitation is
  described by the function $\vect g_e(t)$, the resulting expression
  will be:
  \begin{equation}
    \label{eq:4}
    \vect E(t,Q) = \int_0^{Z_m}N(z)\vect g_e(\tilde t,z)dz,
  \end{equation}
 where $\tilde t$ is the delayed time depending on the geometry of the
system and $N(z)$ describes the dipole density.

The physical nature of the oscillators forming the string can be very
different. In particular, one can use a string of nanoantennas made of
the conducting material. Such nanoantennas have indeed the resonance
frequencies defined by their plasmonic resonances, which are highly
flexible and are determined by the geometry and size of the structures
\cite{novotny11,biagioni12,sun13,feuillet-palma13}.  Using such
structures the resonance frequency $\omega_0$ can be tuned in the wide
range from THz up to the visible. The semiconductor
  quantum dots arrays \cite{hartmann99,felici09,mohan10,juska13} can
  be also used. One may remark that in the case of semiconductors,
  strong nonlinearities accomplished by the possibility of pump allow
  to use such thiner-than-wavelength layers as active elements in
  quantum well and quantum dot lasers
  \cite{iga88,sale:book,chow97,ebeling98,loiko01,babushkin03,babushkin04,schulz-ruhtenberg05,babushkin07a,babushkin08b,schulz-ruhtenberg10}.

If the exciting pulse is weak
  and the dipole response is linear, its response to the excitation pulse $\vect E_e(t)$ is
described by the polarization $\vect P(t)$
  \begin{equation}
    \label{eq:1}
    \ddot{\vect P} + \gamma \dot{\vect P} + \omega_{0}^2 \vect P = g \vect E_e(t),
  \end{equation}
where  $g$ is the coupling strength to the field.

Considering the small signal (linear) regime we
assume also, that the excitation pulse is shorter than the resonant
period of oscillators, so that its spectrum not only includes
$\omega_0$ but at the same time, is significantly broad and flat [see
\reffig{fig1}(c)].  Such pulses can be obtained, for instance, in THz
and MIR range using a gaseous ionization-based source pumped by a
ultrashort optical pump pulse \cite{reimann07, thomson07,kim07,kim08b,
  babushkin10, babushkin10a,koehler11a,babushkin11,berge13}.

Under this condition and also assuming that
  $\gamma\ll\omega_0$, the
nonsingular part of the response of the oscillators can be to a good precision described by an
excitation function
\begin{equation}
  \label{eq:2}
  g_e(t) \approxeq  e^{-\gamma t} \cos(\omega_0 t) \Theta(t),
\end{equation}
where  $\Theta(t)$ is the Heaviside
step-function.

In the case when the signal has large enough
  amplitude, the response of the dipoles becomes nonlinear. The
  nonlinear response in the case of a linear polarization of the
  incident field is described by the Bloch equations
  \cite{Allen:book,kryukov70}:


\begin{eqnarray}
\frac{du(t,z)}{dt}&=&-\Delta\omega v(t,z) - \frac{1}{T_{2}}u(t,z) ,
\label{equ}
\\
\frac{dv(t,z)}{dt}&=&-\Delta\omega u(t,z) - \frac{1}{T_{2}}v(t,z) + \Omega_{R}(t,z)w(t,z) ,~~
\label{eqv}
\\
\frac{dw(t,z)}{dt}&=& - \frac{1}{T_{1}}(w(t,z)+1) - \Omega_{R}(t,z)v(t,z) .~~~~~
\label{eqQ}\end{eqnarray}

Here $u,v$-components of the medium polarization which are in-phase
and out of phase with the driving E field correspondingly, $w$ -
population difference, $T_{1}$ is the time relaxation of the
population difference, $T_{2}$ is the time relaxation of the
polarization, $\Delta\omega$ is the frequency detuning between the
electric field and the  resonance
frequency of the medium, $\Omega_{R}=\frac{d_{12}E(t)}{\hbar}$ - Rabi
frequency of the driving field, $d_{12}$ - transition dipole moment,
$E_{0}$ - amplitude of the driving field. This system of equations
describes interaction of short optical pulses with the two-level
medium, in particular taking into account coherent interaction
\cite{Allen:book,kryukov70}. We remark that
  \refeqs{equ}{eqQ} are derived in the rotating wave approximation,
  that is inapplicable for the broad spectra and short pulses
  mentioned above. Thus, studying the nonlinear response we limit
  ourself with relatively long pump pulses.
Namely we assume in this case that
the excitation pulse has a Gaussian shape given by: $E(t)=E_{0}\exp
\left(\frac{-t^2}{\tau_{p}^2}\right)$. In the case of $\Delta\omega=0$
and if the duration of the excitation pulse is much shorter than
$T_{1}, T_{2}$: $\tau_{p} << T_{1}, T_{2}$ one can obtain the
following solution of the optical Bloch equations for the polarization
$P(t,z)$ and population difference $n(t,z)=n_{0}w(t,z)$ ($n_{0}$ is
the concentration of the two-level atoms):
\begin{eqnarray}
n(t,z)=n_{0}w(t,z)=n_{0}\cos\Phi(t,z), \\
P(t,z)=d_{12}n_{0}u(t,z)=d_{12}n_{0}\sin\Phi(t,z),
\label{eqPN}
\end{eqnarray}
where
\begin{equation}
\Phi(t,z)=\frac{d_{12}}{\hbar} \int_{-\infty}^{t}  E(t',z) dt',
\label{eqPhi}
\end{equation}
is the pulse area \cite{Allen:book,kryukov70}.
The change of $N$ and $P$ can be represented as the rotation
of a unit vector in the $(x, y)$ plane in such a way
that the $x$ component of the vector corresponds to
$v$, and the $y$ component to $w$. Then the function
$\Phi$ is the angle of rotation of this vector: $\Phi = \pi$
corresponds to a complete transition of the particle to
the upper level ($\pi$-pulse), and $\Phi = 2\pi$ corresponds to a complete
return to the ground level ($2\pi$-pulse).

\section{A linear response}

\subsection{Straight string: general considerations}

In this section we consider the case when our string has the form of a
straight line of the length $Z_m$ [\reffig{fig1}(b)].
The observer is located at the very large distance from the string or in the focal point Q of the lens L. The string consists of
identical dipole oscillators having the resonance frequency
$\omega_{0}$ ($\lambda_{0}$ is the corresponding wavelength) and the
decay rate $\gamma$. The number density of oscillators along the Oz
axis varies periodically with the period $\Lambda_{z}$:
\begin{eqnarray}
  N(z)=\frac{1}{2}\left(1+a\cos\frac{2\pi}{\Lambda_{z}}z\right),
\label{eqA}
\end{eqnarray}
where $a\leq1$ is the amplitude of density oscillations. This equation
describes a sort of 1D diffraction grating formed of particles possessing proper resonance frequency. In the following, for
simplicity, we take $a=1$.

At the initial time moment $t=0$ the excitation point
crosses the point $z=0$ and starts to propagate
 at the velocity $V$ along the string
towards its other end.  We suppose that the
exciting pulse is
linearly polarized and the polarization direction is perpendicular to
the plane of Fig.~\ref{fig1}(b).  The oscillators start to emit
electromagnetic radiation according to the response law
\refeqs{eq:2}{eq:3}.  We now consider this
  secondary radiation propagating at the angle $\varphi$ to the
string, which reaches the amplitude detector at the point Q
(Fig.~\ref{fig1}(b)).  Under those circumstances we can restrict
ourself by a single linear polarization, thus obtaining a scalar
problem.

The electric field emitted by the
oscillator located in z (as being
observed in the same point) is proportional, according to
\refeq{eq:2}, to:
\begin{equation}
  E(t,z)
  =\exp\left[-\frac{\gamma}{2}\left(t-\frac{z}{V}\right)\right]\cos\left[\omega_{0}\left(t-\frac{z}{V}\right)\right]\Theta\left[t
    - \frac{z}{V}\right],
\label{eqEz}
\end{equation}
where the delayed argument describes the fact that the excitation
appears in the point $z$ at the moment
delayed by $z/V$. Instead of point Q one may consider
the electric field at the plane S orthogonal to the direction
$\varphi$ passing through the point $Z_m$. Propagation time from this
reference plane to Q is constant and will be omitted in the following
analysis.

The light propagation time from the point $z$ to the reference plane
is given by $\frac{Z_{m}-z}{c}\cos\varphi$. Thus, the electric field
emitted at the point $z$ will have at the reference
plane the value:
\begin{equation}
  E_{ref}(t,z)=\exp\left[-\frac{\gamma}{2} f\left(t,z\right)\right] \cos\left[\omega_{0}f\left(t,z\right)\right]\Theta\left[f\left(t,z\right)\right],
\label{eqEs}
\end{equation}
where $f\left(t,z\right)= t - \frac{z}{V} - \frac{Z_{m}-z}{c}\cos
\varphi$.

The total field observed in $Q$ is obtained  by the integration of \refeq{eqEs} over the whole string:
\begin{gather}
\nonumber
  E(t,\varphi) = \int\limits_0^{Z_{m}} N(z)\exp\left[-\frac{\gamma}{2} f\left(t,z\right)\right]\cos\left[\omega_{0}f\left(t,z\right)\right]\times \\
\times \Theta\left[f\left(t,z\right)\right]\, dz.
\label{eqE}
\end{gather}
The analytical solution of \refeq{eqE} in the case of $\gamma=0$ is given in
the Appendix.  As one can see from analytical calculation in Appendix [see \refeq{eq:a5}]  the response contains
the resonance frequency of oscillators $\omega_{0}$ together with a new
component given, by the
  expression:
\begin{equation}
\Omega_{1} = 2\pi\frac{V/\Lambda_{z}}{|\frac{V}{c}\cos\varphi-1|},
\label{eqOmega}
\end{equation}

The inverse numerator of
  \refeq{eqOmega} is the time interval which the excitation spot needs
  to cross the single oscillation period of $N(z)$.
When this
  time is equal to the period of the oscillations ($V/c=\Lambda_{z}/\lambda_{0}$)
formula (\ref{eqOmega}) leads to:
\begin{equation}
\Omega_{1_{D}} =\frac{\omega_{0}}{|\frac{V}{c}\cos\varphi-1|}.
\label{eqOmega1D}
\end{equation}
This relation formally coincides with that one for the Doppler frequency
shift ~\cite{frank42, bolotovskii72}, so we will call it the Doppler
frequency but will keep in mind that its physical origin differs from that of the Doppler effect.

Equation  \refeq{eqOmega} is valid for  arbitrary
  $V$ and has the same form as the one appearing in the case of
Purcell-Smith radiation. The appearance of this frequency and other related
questions will be studied in detail in the next section.

\subsection{Straight string: the
    linear response dynamics}

Now we explore the temporal and spectral shape of the linear string
response defined by \refeq{eqE} and its dependence on the
system parameters.

We start from the numerical simulations of \refeq{eqE} for some
``typical'' parameter values. Namely, we choose the normalized
parameters as: $\frac{V}{c}=2$, $\frac{Z_{m}}{\Lambda_{z}}=9.55$,
$\frac{\Lambda_{z}}{\lambda_{0}}=5$,
$\frac{\omega_{0}}{\gamma}=22.22$.  The real-world values of the
parameters corresponding to this set depend on the resonance frequency
of the oscillators in the string.  For instance, assuming
$\omega_0=2\pi\times 10$ ps$^{-1}$ (frequency for which the
$\delta$-function assumption from \reffig{fig1}(c) is especially easy
fulfilled), we will
  have $\Lambda_z=150\mu m$, $Z_m=1.4$ mm, and $\gamma=2.8$ ps$^{-1}$. Another example is the pump at optical frequencies
 $\omega_0=2\pi\times 375$
  ps$^{-1}$, $\Lambda_z=4$ $\mu m$, $Z_m=40 \mu m$,
$\gamma=106.04$ ps $^{-1}$.

\begin{figure}[b]
\includegraphics[width=1\linewidth]{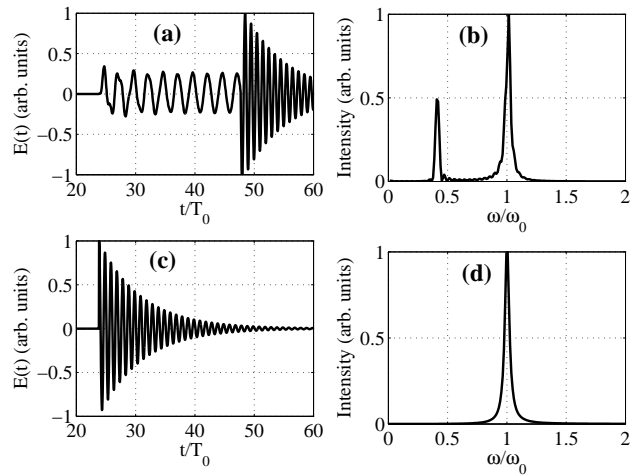}
\caption{Time dependence of the field $E(t)$ according to \refeq{eqE}
  (a,c) and its spectral intensity $I(\omega)$ (b,d) normalized to their maximal
  values vs. normalized time $t/T_0$ and frequency $\omega/\omega_0$,
  for $\frac{V}{c}=2$, $\frac{Z_{m}}{\Lambda_{z}}=9.55$,
  $\frac{\Lambda_{z}}{\lambda_{0}}=5$,
  $\frac{\omega_{0}}{\gamma}=22.22$ for the observation angle $\varphi=0$
  (a,b) and $\varphi=60^\circ$ (c,d), the later corresponds to the
  Cherenkov emission angle.}
\label{fig2}
\end{figure}

The numerical solution of integral \refeq{eqE} and its spectrum
are shown in \reffig{fig2} (in normalized units) for
the above mentioned parameters, and assuming
$\varphi=0$ (observation point is on the same line as the string) in \reffig{fig2}(a,b) and the Cherenkov angle
$\varphi=60$ degree in \reffig{fig2}(c,d).

As one can see, the resonant response at $\omega=\omega_0$ dominates
in both cases. Nevertheless, for $\varphi=0$ additional frequency
arises.  As it is seen in \reffig{fig2}(a), the new frequency appears
in the transient process for the time interval from
approx. $t_{1}/T_{0} =20$ to
approx. $t_{2}/T_{0}=47$, at the moment $t_{1}$
the excitation
spot reaches the end of the
string. During the period  $t_1$ to $t_2$, the radiation from the points $z=Z_m$ to $z=0$
  arrives to the observation plane. As the result of
the interference of the incoming waves, a
 transition process
occurs. It lasts until the moment $t_{2}$. Only the decaying emission with the frequency $\omega_0$ remains at the later time.

 \begin{figure}[t]
\centering
\includegraphics[width=1\linewidth]{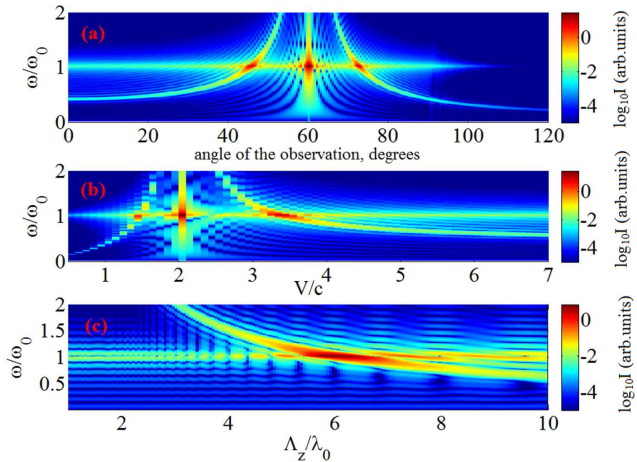}
\caption{Dependence of the spectral intensity $I(\omega)$ of the string response
  according to \refeq{eqE} on the observation angle $\varphi$ (a), the
  excitation velocity $V$ (b) and on the string density
  modulation period $\Lambda_z$ (c). The other parameters coincide
  with those ones given in \reffig{fig2}(c,d).
The spectral intensity is presented in the logarithmic scale.}
\label{fig3}
\end{figure}

For the superluminal velocity of the excitation the denominator of
\refeq{eqOmega} approaches zero if:

\begin{equation}
\cos\varphi_{0} = c/V,
\label{eqCherenkovradcond}
\end{equation}
 which coincides with the condition for Cherenkov radiation. \reffig{fig2}(c) corresponds to the Cherenkov emission angle. This
 angle corresponds also to the zeroes-order diffraction peak of the
 grating formed by $N(z)$. Under the parameters of Fig.~\ref{fig1}, $\varphi_{0} =
 60$ degrees.   When the condition \refeq{eqCherenkovradcond} is
 fulfilled, we have $\Omega_{1}=\infty$, and the radiation from
 all points of the grating (the resonance medium) comes to the
 reference plane simultaneously, thus no transient process occurs.

Analogously, +1st- and -1st diffraction orders maxima are defined by
the relation:
\begin{equation}
\cos\varphi_{\pm1} = \frac{\pm\lambda_{0}}{\Lambda_{z}} + \frac{c}{V},
\label{eqplusminus1}
\end{equation}
which for the parameters of Fig.~\ref{fig3}(a) gives the angles
$\varphi_{+1} = 45.57$ and $\varphi_{-1} = 72.54$
degree correspondingly. For those angles, we have
$\omega_{0} = \Omega_{1}$ as well.  For all
   values of $\varphi$ different from the one given by \refeq{eqplusminus1}, the Doppler
  frequency $\Omega_1$ is not equal to $\omega_0$. It should be
noted however that in this
  case the radiation intensity is smaller than for the Cherenkov
angle.

Dependence of the \refeq{eqE} solution spectrum on the
system parameters is presented in \reffig{fig3}. In
particular, the dependence on the observation angle $\varphi$ is
presented in Fig.~\ref{fig3}(a), on the excitation speed $V$ in
Fig.~\ref{fig3}(b) and on the grating period $\Lambda_z$
(cf. \refeq{eqA}) in Fig.~\ref{fig3}(c).

Dependence of the string response spectrum on $V/c$ assuming
$\varphi=60$ degree is presented in Fig. ~\ref{fig3}(b). The other
parameters are the same as in Fig.~\ref{fig2}(c). One can clearly see
the frequency branch corresponding to the resonance $\omega=\omega_0$,
as well as the another
one corresponding to the  frequency
shift given by \refeq{eqOmega}.

According to Eq.~(\ref{eqOmega}) and
Fig.~\ref{fig2}(c), $\Omega_1$ decreases with the
increasing of $V/c$ for $V/c > 2$ and
increases for $V/c < 2$. From Eq.~(\ref{eqOmega}) it
also follows that $\Omega_{1}\to\infty$ for $V\to 2c$ (when
$\varphi=60$ degree).
This also coincides with the typical behavior of the Doppler frequency
shift.

The dependence of the string
response spectrum on the modulation period
$\Lambda_{z}/\lambda_{0}$ is presented in Fig.~\ref{fig3}(c) for $V/c=3, \varphi=60$ degree. As it
can be seen, $\Omega_{1}$ decreases with increasing of the
$\Lambda_{z}/\lambda_{0}$.


Up to now we have considered the
case when the string is excited by a
 spot of light moving at the
  superluminal velocity. Another
interesting case if the  exciting spot moves at the sub-luminal
velocity.

Such situation can be realized not only
  using the scheme in \reffig{fig1}, but also using an electron beam
moving with some velocity $u$ at an angle $\psi$ to
the boundary of the string. In this case, the velocity of the intersection
of the incident beam with the boundary of the medium is $V=u/\sin\psi$ \cite{bolotovskii72}.

\begin{figure}[h]
\center{\includegraphics[width=0.9\linewidth]{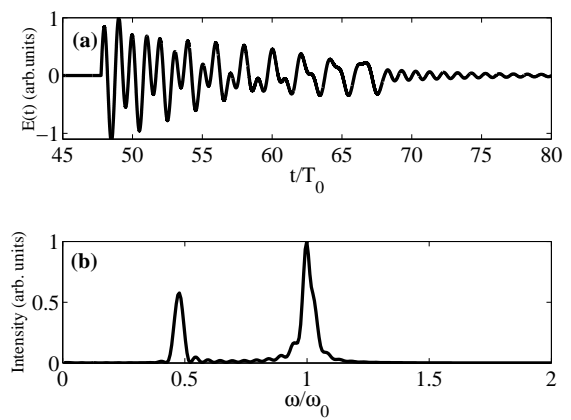}}
\caption{ Time dependence (a) of the
  string response field $E(t)$ according to
 \refeq{eqE}, and (b)
  its spectral intensity $I(\omega)$ normalized to
  the maximum values vs. normalized time $t/T_0$ and
  frequency $\omega/\omega_0$, for $\frac{V}{c}=0.7$,
  $\frac{Z_{m}}{\Lambda_{z}}=9.55$,
  $\frac{\Lambda_{z}}{\lambda_{0}}=5$,
  $\frac{\omega_{0}}{\gamma}=22.22$
  and observation angle $\varphi=0$.}
\label{fig1a}
\end{figure}

The example of numerical solution of the integral
\refeq{eqE} assuming $V/c=0.7$ and $\varphi=0$ are shown in
\reffig{fig1a}.  Other parameters are the same as in
the \reffig{fig2}.  One can see that the
additional frequency component arises during the transient process from approx. $t_{1}/T_{0}=47$ to
approx. $t_{2}/T_{0}=70$. At the time moment $t_{1}$ the radiation from the
point $z=0$ reaches the end of the medium. At the time moment $t_{2}$
the radiation from the point $z=Z_{m}$ appears at the observation
point. Later on, only decaying
oscillations at  the
frequency $\omega_{0}$ remain.

\begin{figure}[h]
\center{\includegraphics[width=1.0\linewidth]{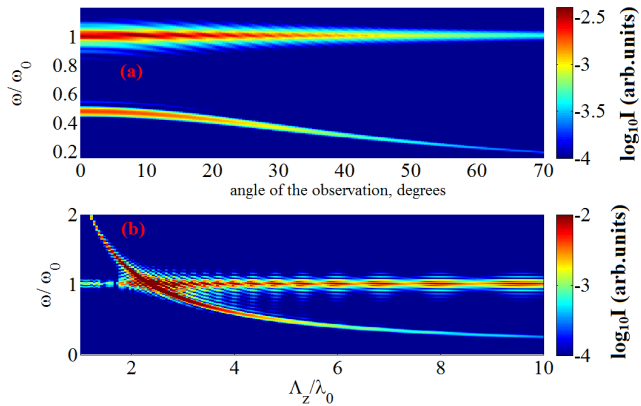}}
\caption{Dependence of the spectral intensity $I(\omega)$ of the
  string response according to
  \refeq{eqE} on the observation angle $\varphi$ (a),
  and the string density modulation period $\Lambda_z$ (b). The other
  parameters coincide with ones in
 \reffig{fig1a}. Note the logarithmic scale
  in the plot.}
\label{fig2a}
\end{figure}

Dependence of the string response spectrum  on the observation angle $\varphi$ as
  well as on the grating period $\Lambda_z$  is
presented in \reffig{fig2a}. As one can see, the
  situation in the case $V<c$ is in many respects similar to the
  case of the superluminal velocity. In
  particular, $\Omega_{1}$ decreases with the increase of
 $\varphi$ as well as with the increase of
$\Lambda_{z}$. On the other hand, the Cherenkov angle at which
  $\Omega_1=\infty$ is never achieved.




\subsection{Circular string: general considerations}

\begin{figure}[tp]
\centering
\center{\includegraphics[width=0.5\linewidth]{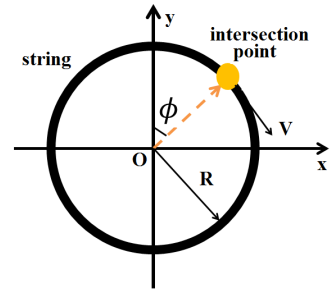}}
\caption{Circular geometry of the string. The source of a short pulse
  with a broad spectrum (c) is located in the
  center of the circle and quickly rotates. The cross-section of the
  pulse an medium (yellow dot) moves at the velocity $v$ along the
  string (black circle). As in the previous case, the string is made
  of dipoles characterized with resonance frequency $\omega_0$ and the dipoles
  number density is modulated along the string periodically with the angular
  period $\Lambda_\phi$.}
\label{fig4}
\end{figure}

In this section we consider completely different topology of the
string depictured in Fig.~\ref{fig4}. Namely, the string made of the
dipoles owing the same resonance frequency $\omega_0$ as before is
arranged along the circle of radius
$R$. The dipole density is modulated along the string in a periodical
way with the angular period $\Lambda_\phi$ as:
\begin{equation}
N(\phi)=\frac{1}{2}\left(1+a \cos(\frac{2\pi}{\Lambda_{\phi}}\phi)\right).
\label{eqB}
\end{equation}
 As in \refeq{eqA}, we assume the modulation amplitude $a=1$.
In the center of the circle a source of a short spectrally broad
optical pulse (see Fig.~\ref{fig1}(c)) is located, which quickly
rotates, so that the cross-section point (yellow point in Fig.~\ref{fig4})
moves at the velocity $V$ along the circle.

The dipoles of the string response to the excitation emitting  the secondary waves. Here we will concentrate on the behavior of the string
response field $E(t)$ observed in the center of circle.  The electric field formed in the center of the circle by
the element $dE_{\phi}$ located at the point which angular coordinate
is $\phi$ is given by:
 \begin{eqnarray}
\nonumber
dE_{\phi}(t)=N(\phi)\exp\left[-\frac{\gamma}{2}f_{\phi}\left(t,\phi\right)\right]\cos\left[\omega_{0}f_{\phi}\left(t,\phi\right)\right]\times \\
\times\Theta\left[f_{\phi}\left(t,\phi\right)\right]d\phi,
\label{eqEphi}
\end{eqnarray}
where $f_{\phi}\left(t,\phi\right)=t - \frac{R\phi}{V} - \frac{R}{c}$.
For one round pass of the excitation, the total electric field is obtained by integration (\ref{eqEphi})
over $\phi$:
\begin{gather}
\nonumber
  E(t,\phi) = \int\limits_0^{2\pi} N(\phi)\exp\left[-\frac{\gamma}{2}f_{\phi}\left(t,\phi\right)\right]\cos\left[\omega_{0}f_{\phi}\left(t,\phi\right)\right]\times \\
\times\Theta\left[f_{\phi}\left(t,\phi\right)\right]d\phi.
\label{Ecent}
\end{gather}

The analytical solution of Eq.~(\ref{Ecent}) in the case of $\gamma=0$ is given in the Appendix. As one can see from analytical calculation in Appendix [see \refeq{eq:a8}]  the response contains
the resonance frequency of oscillators $\omega_{0}$ together with a new
component given, by the
  expression:

\begin{equation}
\Omega_{2} = 2\pi\frac{V/\Lambda_{\phi}}{R}.
\label{eqOmega2}
\end{equation}
 Eq.~(\ref{eqOmega2}) also has a simple physical meaning, namely this
 is the frequency at which the intersection point crosses the
 inhomogeneity oscillations.  Under the condition
\begin{equation}
\frac{V}{c}=\frac{\Lambda_{\phi}R}{\lambda_{0}},
\label{eqres2}
\end{equation}
 the new frequency is equal to the resonance one.

Note also that \refeq{eqOmega2} is valid if the observer is
located anywhere on the axis passing through the center of the circle
perpendicularly to its plane.

\subsection{Circular string: the linear response dynamics}

\begin{figure}[tph!]
\centering
\center{\includegraphics[width=1\linewidth]{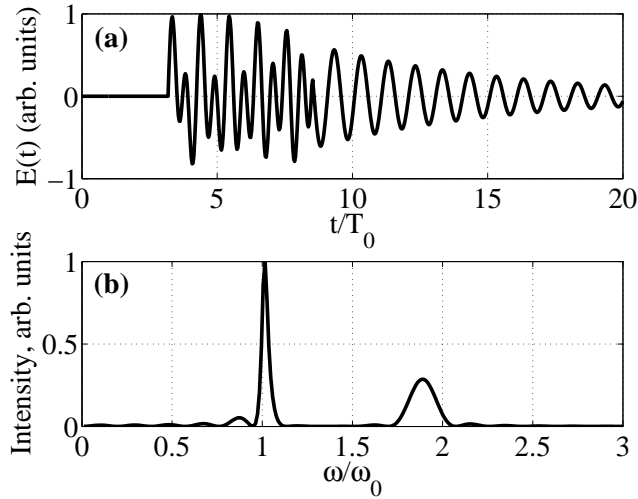}}
\caption{(a) Time dependence of the electric field $E(t)$ excited by
  the string and (b) the corresponding intensity spectrum $I(\omega)$
  in the center of the circle for the circular scheme depictured in
  \reffig{fig4} and the parameters $V/c=3.75$,
  $\frac{\Lambda_{\phi}R}{\lambda_{0}}=2$, $\omega_{0}/\gamma=22.2$,
  $\omega_{0}/\Omega_{2}=0.53$.}
\label{fig5}
\end{figure}

We start from a typical situation in the spectrum when the frequency
$\Omega_2$ is clearly visible. Namely, we take the following
parameters: $V/c=3.75$, $\frac{\Lambda_{\phi}R}{\lambda_{0}}=2$,
$\omega_{0}/\gamma=22.2$, $\omega_{0}/\Omega_{2}=0.53$.
Assuming the same $\omega_0=2\pi\times 10$ ps$^{-1}$
  as in the Sec. III B and $R=3$ cm, we obtain $\Lambda_{\phi}=0.002$ rad$^{-1}$,
$\gamma=2.8$ ps$^{-1}$.
The transient process for these parameters calculated using
\refeq{Ecent} is shown  in \reffig{fig5}.
For these particular parameters,
the frequency of oscillators practically doubles that of the transient
emission, it results in the high-amplitude beatings clearly seen in
\reffig{fig5}. Once the transition process is finished,
the observer at O records the ordinary decaying oscillations.  This conclusion is also valid in the case when the
excitation pulse moves at the sub-luminal velocity or precisely at
the velocity of light. In all these cases, the radiation spectrum at
the center of circle will possess a new frequency, with only exception
of the resonance \refeq{eqres2}.

\begin{figure}[tp]
\centering
\center{\includegraphics[width=1\linewidth]{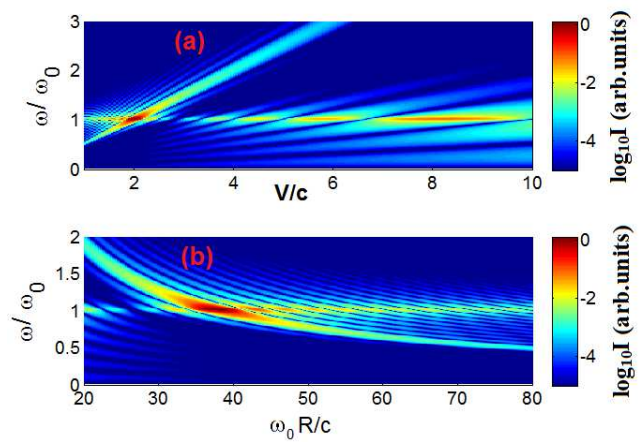}}
\caption{(a) Dependence of the radiation spectrum on the normalized
  propagation speed $V$ of the excitation (a) and the radius of the
  circle $R$ (b). Other parameters are as in the Fig. ~\ref{fig5}}
\label{fig6}
\end{figure}

In order to illustrate the dependence $\Omega_{2}$ on the parameters
of system, we present the radiation spectrum in dependence on $V$
(\reffig{fig6}(a)) and $R$ (\reffig{fig6}(b)) whereas the other
parameters are taken as in \reffig{fig5}.  As it can be easily
seen from \refeq{eqOmega2}, the new frequency increases with the
increase of $V$ and decreases with  $R$.

In the presented circular case, the role of the angle (if the
excitation velocity is fixed) plays the radius of the circle. The Cherenkov
resonance corresponds then to the R value defined by \refeq{eqres2}.

\section{Strong pump and  nonlinear
 dynamics}

 In
the present section we
investigate a response of
a straight string
in the regime when the pump is strong enough, so that
  the response of the string is significantly nonlinear.

The response of the string, assuming the scalar
  approximation and also taking into account that the secondary
  radiation never comes back to the string and thus no nonlinear
  propagation takes place (unless the observation angle $\varphi=0$) is
  described by \refeq{eq:4}. Taking into account \refeq{eqPN} we
  obtain:
\begin{gather}
  E(t,\varphi) = \int\limits_0^{Z_{m}} N(z)P\left[f(t,z)\right]\sin\left[\omega_{0}f\left(t,z\right)\right]dz.
\label{eqEnn}
\end{gather}

\begin{figure}[tph!]
\centering
\center{\includegraphics[width=1\linewidth]{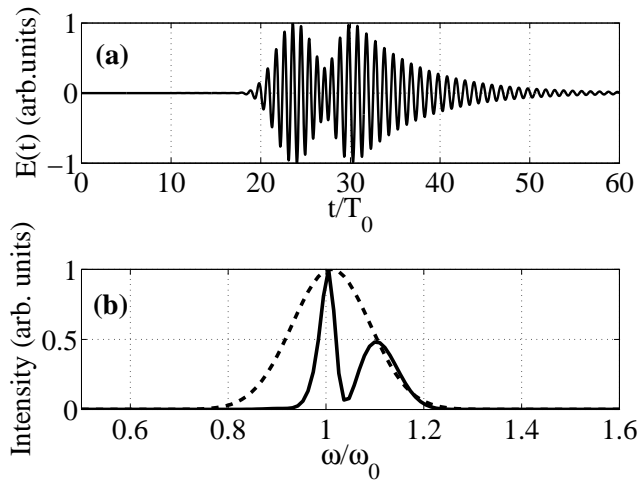}}
\caption{Time dependence (a) of the string response field $E(t)$
  according to \refeq{eqEnn}, and (b) solid line - its spectral
  intensity $I(\omega)$ normalized to the maximum values
  vs. normalized time $t/T_0$ and frequency $\omega/\omega_0$, for
  the parameters of \reffig{fig2} and total
  pulse area $S=\pi/2$, $\Omega_{R}=0.07\omega_{0}$,
  $\tau_{p}=2T_{0}$, dashed line - spectrum of the excitation pulse}
\label{fig7}
\end{figure}

In this section, as it was already mentioned, we
  consider pulses with relatively narrow spectrum, to be consistent
  with the approximations for which \refeq{eqPN} were derived. The
  pulses we use are nevertheless still short enough to clearly observe
  the frequency $\Omega_1$. The result of numerical solution of the
integral \refeq{eqEnn} assuming the parameters of
  \reffig{fig2} and the observation angle $\varphi=71$ degree, total pulse area
$S=\pi/2$, $\Omega_{R}=0.07\omega_{0}$ and $\tau_{p}=2T_{0}$ is
presented in the \reffig{fig7}.

As one can see from the \reffig{fig7}(a), analogously
  to the linear case, the system
demonstrates a short pulse in a transient
regime. Its duration is equal to few
periods of optical oscillations $T_{0}$. After a
transient process decaying emission
at the frequency $\omega_{0}$ is observed.
As in the linear case, the two frequencies are
  observed as shown in \reffig{fig7}(b). One may note that the peak
  corresponding to the frequency $\Omega_1$ is more pronounced than
  in the linear case.

\begin{figure}[tp]
\centering
\center{\includegraphics[width=1\linewidth]{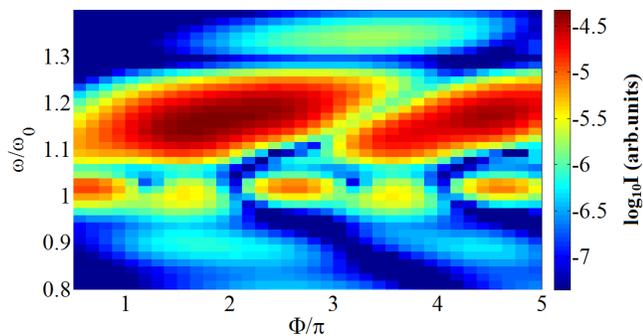}}
\caption{Dependence of the radiation spectrum on the pulse area $\Phi$.
  Other parameters are as in the Fig. ~\ref{fig7}}
\label{fig8}
\end{figure}

\reffig{fig8} illustrates the dependence of the
secondary emission spectrum on the total pulse area
$\Phi$ (cf. \refeq{eqPhi}). Pulse area was changed via modification of the Rabi frequency
(pulse amplitude), keeping the
pulse duration constant. From \reffig{fig8} one can
observe two branches
corresponding to the resonance frequency $\omega_{0}$ and
to the Doppler-like one
$\Omega_{1}$.  Analysis of \reffig{fig8} shows that in
strongly nonlinear regime when the pulse area is
large the radiation on the resonance frequency $\omega_{0}$ has
even smaller intensity than the one on  the frequency $\Omega_{1}$. The periodic structure revealed in \reffig{fig8} in
  dependence on $\Phi$ is explained by phase relations between the
  periodic term $\sin\Phi$ entering $P(t,z)$
  [cf. \refeq{eqEnn},~\refeq{eqPN} and \refeq{eqPhi}] and the period of
  spatial inhomogeneity of dipole density $N(z)$.

\section{Conclusion}
In this paper, the secondary radiation excited by a moving
intersection of a short spectrally broadband pulse and a resonant
string made of identical dipoles is discussed for the linear and
circular string geometry.  In such a situation, the Cherenkov radiation
  naturally appears. In contrast to many other cases where Cherenkov
radiation is unstructured and has no clear frequency resonance, the
present one demonstrates obvious resonant properties. That is, the
response spectrum is centered at the resonant frequency of the dipoles
comprising the string.
In addition, as our analysis shows, a new frequency appears in the
presence of the string density oscillations, which has the meaning of
a Doppler shift of the resonant frequency $\omega_0$.

We point out also that the new Doppler frequency ($\Omega_1$ in the
case of the straight string and
$\Omega_2$ in the circular case) appears in the transient regime, when
some of the secondary waves excited by the
exciting pulse have not yet reached the observation
plane.  The dynamics of the radiation after this moment is trivial and
contains only decaying oscillations on the resonant frequency
$\omega_0$. In the strong-signal regime, when the nonlinearity in the
string response becomes significant, the new frequency may even
significantly overcome the resonant one.

The behavior described there can find its application, for instance,
to shape the broad spectra and short pulses in desired way using rather
compact setup.

\begin{acknowledgments}
R.M. Arkhipov would like to acknowledge the support of EU FP7 ITN
PROPHET, Grant No. 264687. I. B. is thankful to German
  Research Foundation (DFG) for the financial support in the framework
  of the Collaborative Research Center SFB 910 and project BA 41561-1.

\end{acknowledgments}

\appendix

\section{\label{app:1} Analytical solutions of \refeq{eqE} and \refeq{Ecent}.}

In this appendix
  we provide an analytical solutions of
\refeq{eqE} and \refeq{Ecent} when $\gamma=0$.
To obtain such expression we first rearrange the argument of $\Theta$
- function in \refeq{eqE} as $f\left(t,z\right)=t - \frac{z}{V} -
\frac{Z_{m}-z}{c}\cos \varphi = t - z/W - \frac{Z_{m}}{c}\cos\varphi$,
where $W$ is the effective velocity defined as:
\begin{equation}
\frac{1}{W} = \frac{1}{V} - \frac{1}{c/\cos\varphi}.
\label{eqA1}
\end{equation}

One can see that $W$ can be interpreted as the velocity of the
projection of the cross-section point to the axis parallel to the
observation plane in \reffig{fig1}(b).
Using this parameter we can rewrite the integral for the pulse
response in the form:
\begin{equation}
E(t) = \int N(z)h_{0}\left(t' - \frac{z}{W}\right) dz,
\label{eqA2}
\end{equation} 
where $t' = t - \frac{Z_{m}\cos\varphi}{c}$,
and the function $h_{0}(t)$ denotes the response of a dipole located
at $z=0$ and excited with the excitation in the form of delta-function
$\delta(t)$: $h_{0}(t)=\cos(\omega_{0}t)\Theta(t)$.

The integral in \refeq{eqA2} has different form depending on the sign of
$W$: 
\begin{eqnarray}
E(t) = \int_0^{Wt'}N(z)h_{0}\left(t' - \frac{z}{W}\right)dz
\,\mathrm{for}\, W>0,
\label{eqE1}
\\
E(t) = \int_{Z_{m}}^{Wt'}N(z)h_{0}\left(t' - \frac{z}{W}\right)dz \,\mathrm{for}\, W<0.
\label{eqE2}
\end{eqnarray}
If $W>0$ the emitting element of the string
moves in positive direction starting from zero as being seen by the
observer. In the opposite situation, when $W<0$, it is seen as moving in the negative
direction from $Z_{m}$ to 0.

\refeqs{eqE1}{eqE2} are valid for
$0<t<Z_{m}/|W|$ (transient regime)  assuming
$\nu_{z} = \frac{2\pi}{\Lambda_{z}}$.
  In particular for $W>0$ one can obtain:
\begin{gather}
\nonumber E(t)=
 \frac{W}{\omega_{0}}\sin\left(\omega_{0}t'\right) + \\
\frac{W}{W^2\nu_{z}^2 -\omega_{0}^2}\left[
  \nu_{z}W\sin\left(\nu_{z}Wt'\right) -
  \omega_{0}\sin\left(\omega_{0}t'\right) \right].
\end{gather}
On the other hand, for $W<0$ we have:
\begin{gather}
\nonumber
E(t)= \frac{W}{\omega_{0}}\sin\left(\omega_{0}(t'-\frac{Z_{m}}{W})\right)  + \\
\nonumber
+\frac{W^2\nu_{z}\sin(\nu_{z}Wt')
 -\nu_{z}W^2\sin(\nu_{z}Z_{m})\cos\left[\omega_{0}(t'-\frac{Z_{m}}{W})\right]}{{\nu_{z}^2W^2-\omega_{0}^2}} -\\
-\frac{\omega_{0}W\cos(\nu_{z}Z_{m})\sin\left[\omega_{0}(t'-\frac{Z_{m}}{W})\right]}{{\nu_{z}^2W^2-\omega_{0}^2}}.
\label{eq:a5}
\end{gather}

The last equations contain the oscillating terms with the frequencies $\omega_{0}$ and $\Omega_{1}=\nu_{z}W$ which coincides with the \refeq{eqOmega}.

For $t\geq Z_{m}/W$, that is when the excitation pulse comes out of
the string, we have:
\begin{gather}
\nonumber
E(t)= \int_0^{Z_{m}}N(z)\cos\left[\omega_{0}\left(t' - \frac{z}{W}\right)\right]dz =\\ \nonumber
=\frac{W}{\omega_{0}}\left[\sin(\omega_{0}t')-\sin\left(\omega_{0}(t'-\frac{Z_{m}}{W})\right)\right]  + \\
\nonumber
\frac{\nu_{z}W^2\sin(\nu_{z}Z_{m})\cos\left[\omega_{0}(t'-\frac{Z_{m}}{W})\right]}{{\nu_{z}^2W^2-\omega_{0}^2}} + \\
+\frac{\omega_{0}W\cos(\nu_{z}Z_{m})\sin\left[\omega_{0}(t'-\frac{Z_{m}}{W})\right]-W\omega_{0}\sin(\omega_{0}t')}{{\nu_{z}^2W^2-\omega_{0}^2}}.
\end{gather}
This term describes the oscillations with the frequency $\omega_{0}$ after transition process stops.

In the case of circular geometry for transient process ($\frac{R}{c}<t<\frac{2\pi R}{V} + \frac{R}{c}$ if $V>c$) one can obtain

\begin{widetext}
\begin{gather}
\nonumber
E(t)= \int_{Vt''/R}^{2\pi}N(\phi)\cos\left[\omega_{0}\left(t'' - \frac{R\phi}{V}\right)\right]d\phi 
=-\frac{V}{R\omega_{0}}\sin\left(\omega_{0}(t''-\frac{2\pi R}{V})\right)  + \\ 
+\frac{V^2\nu_{\phi}\sin(2\pi\nu_{\phi})\cos\left[\omega_{0}(t''-\frac{2\pi R}{V})\right]}{{\nu_{\phi}^2V^2-R^2\omega_{0}^2}} 
+\frac{R\omega_{0}V\cos(2\pi\nu_{\phi})\sin\left[\omega_{0}(t''-\frac{2\pi R}{V})\right]}{{\nu_{\phi}^2V^2-R^2\omega_{0}^2}} 
  -\frac{V^2\nu_{\phi}\sin(\Omega_{2}t'')}{{\nu_{\phi}^2V^2-R^2\omega_{0}^2}}.
\label{eq:a8}
\end{gather}
\end{widetext}
Here $t''=t-\frac{R}{c}$, $\nu_{\phi}=\frac{2\pi}{\Lambda_{\phi}}$. The last expression contains terms oscillating on the frequencies $\Omega_{2}$ and $\omega_{0}$. After transition process ends ($V>c$, $t>\frac{2\pi R}{V} + \frac{R}{c}$), we have:
\begin{widetext}
\begin{gather}
\nonumber
E(t)= \int_{0}^{2 \pi}N(\phi)\cos\left[\omega_{0}\left(t'' - \frac{R\phi}{V}\right)\right]d\phi 
=\frac{V}{R\omega_{0}}\left[\sin\left(\omega_{0}t''\right)-\sin\left(\omega_{0}(t''-\frac{2\pi R}{V})\right)\right]  + \\ 
+\frac{V^2\nu_{\phi}\sin(2\pi\nu_{\phi})\cos\left[\omega_{0}(t''-\frac{2\pi R}{V})\right]}{{\nu_{\phi}^2V^2-R^2\omega_{0}^2}} 
+\frac{R\omega_{0}V\left[\cos(2\pi\nu_{\phi})\sin\left[\omega_{0}(t''-\frac{2\pi R}{V})\right]-\sin\left(\omega_{0}t''\right)\right]}{{\nu_{\phi}^2V^2-R^2\omega_{0}^2}}.
\end{gather}
\end{widetext}
which contains only terms oscillating with the frequency $\omega_{0}$.


%

\end{document}